\documentclass[reprint,twocolumn,showpacs,amsmath,amssymb,superscriptaddress,aps,prl]{revtex4-1}

\usepackage{graphicx}
\usepackage[english]{babel}
\usepackage[hidelinks]{hyperref}

\newcommand{\RR}{\right}
\newcommand{\LL}{\left}
\newcommand{\m}{\mathrm}
\newcommand{\dg}{\dagger}

\newcommand{\fref}[1]{Fig.~\ref{#1}}

\begin{document}

\title{Quantum back-action evading measurement of collective mechanical modes}

\author{C. F. Ockeloen-Korppi}
\author{E. Damsk\"agg}
\author{J.-M. Pirkkalainen}
\affiliation{Department of Physics, Aalto University, P.O. Box 15100, FI-00076 AALTO, Finland}
\author{A. A. Clerk}
\affiliation{Department of Physics, McGill University, 3600 rue University, Montr\'{e}al, Quebec H3A 2T8, Canada}
\author{M. J. Woolley}
\affiliation{School of Engineering and Information Technology, UNSW Canberra, ACT, 2600, Australia}
\author{M. A. Sillanp\"a\"a}
\email[]{mika.sillanpaa@aalto.fi}
\affiliation{Department of Physics, Aalto University, P.O. Box 15100, FI-00076 AALTO, Finland}

\date{\today}

\begin{abstract}
The standard quantum limit constrains the precision of an oscillator position measurement. It arises from a balance between the imprecision and the quantum back-action of the measurement. However, a measurement of only a single quadrature of the oscillator can evade the back-action and be made with arbitrary precision. Here we demonstrate quantum back-action evading measurements of a collective quadrature of two mechanical oscillators, both coupled to a common microwave cavity. The work allows for quantum state tomography of two mechanical oscillators, and provides a foundation for macroscopic mechanical entanglement and force sensing beyond conventional quantum limits.
\end{abstract}

\maketitle

The interplay of measurement imprecision and quantum back-action limits the sensitivity with which the position of an oscillator can be continuously monitored, to be at best equal to the oscillator's zero-point fluctuations \cite{Caves1980,GirvinReview}. This is known as the standard quantum limit (SQL). However, any single quadrature of the motion can in principle be measured without limit, provided that the measurement back-action is shunted to the orthogonal quadrature. Such back-action evading (BAE) measurements \cite{Caves1980,BraginskyQND,Marquardt2008Sq}, which are examples of quantum non-demolition (QND) measurements, can be achieved by appropriately synchronizing the measurement with the oscillator's intrinsic motion. Classical analogs of BAE measurements have been demonstrated a long time ago \cite{Visco1996BAE,Heidmann2007BAE,Schwab2010QND}. In cavity optomechanical systems, where a mechanical oscillator is dispersively coupled to a driven optical or microwave cavity, BAE measurements that evade the quantum back-action \cite{Lehnert2009SQL,Atom2012Sq,RegalShot} have recently been demonstrated \cite{Schwab2014QND}. In the context of generating squeezed states of mechanical motion \cite{SchwabSqueeze,Squeeze,Teufel2015Squ} they have been used for detection \cite{SchwabSqueeze,Teufel2015Squ}, and they have also been demonstrated in atomic spin systems \cite{Polzik2015Sq}.

\begin{figure}[!ht]
\includegraphics[width=\columnwidth]{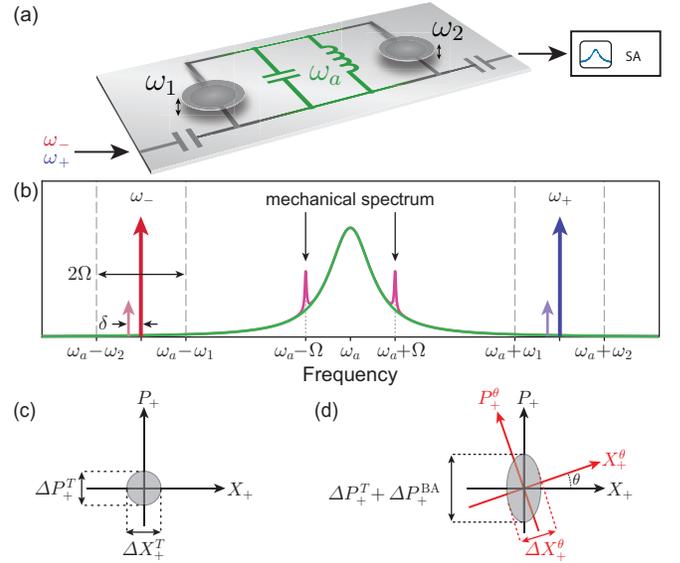}
\caption{\emph{Two-mode BAE measurement scheme.} (a) Schematic representation of our setup. Two micromechanical oscillators ($\omega_1$, $\omega_2$) are capacitively coupled to a superconducting microwave resonator ($\omega_a$). Pump tones are injected to port 1, and the output spectrum of port 2 is measured on a signal analyzer (SA). (b) Two strong microwave pump tones (frequencies $\omega_{-}$ and $\omega_{+}$) realize a BAE measurement. The mechanical spectrum appears as sidebands (pink peaks) on the thermal cavity spectrum (green line). The back-action can be probed by a set of weak probe tones (short arrows), slightly detuned from the pump tones. (c) In thermal equilibrium, the fluctuations (gray circle) of all collective quadratures are equal to $\Delta X_+^T$. (d) The back-action of the BAE measurement heats up the $P_+$ quadrature, whereas $X_+$ remains unaffected. The red axes represent the projection on a basis defined by the probes.}
\label{fig:setup}
\end{figure}

BAE techniques have previously been discussed for two coupled mechanical oscillators \cite{Rioli1993} and optomechanical systems \cite{TsangCaves2010,TsangCaves2012}, and demonstrated for two atomic spin ensembles \cite{WasilewskiPolzik2010}. In a recent theoretical work \cite{WoolleyBAE}, the concept has been extended to the collective modes of two uncoupled mechanical oscillators, each independently coupled to an electromagnetic cavity. This type of measurement allows one to measure both quadratures of a narrow-band force applied to one of the oscillators without any fundamental quantum limit \cite{WoolleyBAE}. Adding feedback control \cite{WoolleyBAE}, or perturbing the measurement slightly (i.e., reservoir engineering) \cite{ClerkEnt2014}, such measurements could be used to generate steady-state entanglement between two macroscopic mechanical oscillators.

In this Letter, we for the first time experimentally demonstrate such a mechanical two-mode BAE measurement. We simultaneously achieve a measurement imprecision below the quantum zero-point fluctuations and an evasion of quantum back-action caused by microwave shot noise (below the back-action arising in a continuous position measurement). The canonically conjugate quadrature is heated predominantly by the quantum back-action.

Our system is shown in \fref{fig:setup}(a). It consists of a microwave cavity resonator and two mechanical oscillators which have no direct coupling. Previously, optomechanical systems containing more than one mechanical oscillator have been experimentally studied both in the optical \cite{PainterMix2010,Lipson2012sync,Harris2014TwoMode,TwoMode2014,Stamper-Kurn2016} and microwave \cite{multimode2012,Weig2012LZ} regimes. Our cavity is a superconducting on-chip $LC$ resonator, with frequency $\omega_a$, decay rate $\kappa$, and mode operator $a$. The mechanical oscillators are realized as aluminum drumheads with mode operators $b_i$, frequencies $\omega_i$, and decay rates $\gamma_i$ ($i=1,2$). Each mechanical oscillator is individually coupled to the cavity via the radiation-pressure interaction $H_i=g_ia^\dg a\LL(b_i^\dg+b_i\RR)$, realizing a three-mode cavity optomechanical system \cite{OptoReview2014}. The single-photon coupling rates $g_i$ arise from a position-dependent capacitance between the cavity and each mechanical oscillator. The total Hamiltonian is hence $H=\omega_a a^\dg a+\sum_{i}\LL(\omega_ib_i^\dg b_i+H_i\RR)$. We describe the system in a reference frame set by the cavity frequency and the average of the two mechanical frequencies \cite{supplement}. In this frame we define the position quadrature $X_i=(b^\dg_i+b_i)/\sqrt{2}$ and momentum quadrature $P_i=i(b^\dg_i-b_i)/\sqrt{2}$, and then the collective quadrature coordinates $X_\pm=\LL(X_1\pm X_2\RR)/\sqrt{2}$ and $P_\pm=\LL(P_1\pm P_2\RR)/\sqrt{2}$. Note that these quadratures are not QND variables, but have (as we will see) useful dynamics. When each oscillator is in equilibrium with a bath of temperature $T_i$, it has the thermal occupation $n^T_i\simeq k_\m{B}T_i/\hbar\omega_i$. In such a thermal state, the variances of all collective quadratures are equal: $\LL(\Delta X_\pm^T\RR)^2=\LL(\Delta P_\pm^T\RR)^2=\LL(n^T_1+n^T_2+1\RR)/2$, as depicted in \fref{fig:setup}(c).

\begin{figure}
\includegraphics[width=\columnwidth]{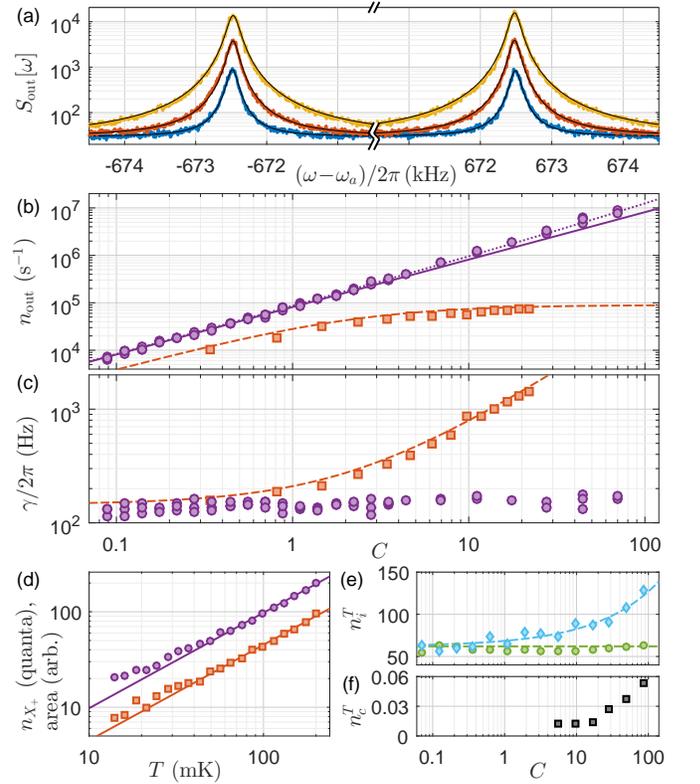}
\caption{\emph{Back-action evading measurement.} (a) Cavity output spectra at $\omega_a-|\Omega|$ (left) and at $\omega_a+|\Omega|$ (right), providing a measurement of the $X_+$ collective quadrature, for $C=4.4,18,70$ (bottom to top). The solid lines are Lorentzian fits. (b) Integrated area of the spectra (purple dots) as a function of measurement strength given by the cooperativity. The solid line is expected for perfect BAE; the dotted line includes technical heating. The red squares correspond to cooling \cite{supplement}. (c) Effective linewidth of the spectral peaks corresponding to the data in (b). (d) BAE thermometry (purple dots, $n_{X_+}$) for $C=0.2$ as function of cryostat temperature. Solid line is a fit to the data, used to calibrate the mechanical signal. Red squares show the integrated area for cooling (arbitrary units). (e,f) Effective thermal occupation of the (e) mechanical oscillators and (f) cavity, measured with a single red-sideband pump. Mechanical oscillator 2 (blue diamonds) shows technical heating. Dashed lines correspond to an empirical heating model.}
\label{fig:bae_power}
\end{figure}

In order to achieve two-mode BAE using only two tones, the system is simultaneously pumped at the frequencies $\omega_{\pm}=\omega_a\pm(\omega_1+\omega_2)/2$ with equal (real) amplitudes $\bar{a}$, as shown in \fref{fig:setup}(b) \cite{WoolleyBAE}. Assuming the system is in the resolved-sideband limit $\omega_i\gg\kappa$, we neglect terms oscillating at $\pm 2\omega_i$. Further assuming that $g_1 \approx g_2$, the effective Hamiltonian is
\begin{equation}
\label{eq:hamilt}
\begin{split}
& H = \Omega \LL(X_+ X_- + P_+ P_- \RR) + 2 G \LL(a^\dg + a  \RR)  X_+ \,,
 \end{split}
 \end{equation}
where $\Omega=(\omega_1-\omega_2)/2$ is the effective mechanical oscillator frequency, and the effective optomechanical coupling is $G=(g_1+g_2)\bar{a}/2$ \cite{supplement}. Under this Hamiltonian, $X_+$ and $P_-$ act dynamically like the position and momentum of a single harmonic oscillator, even though they commute with one another \cite{HammererZoller2009,TsangCaves2012,WoolleyBAE}. It is an example of the recently introduced concept of a ``quantum-mechanics-free subsystem'' \cite{TsangCaves2012}, previously observed in an atomic system~\cite{WasilewskiPolzik2010}. Only the $X_+$ collective quadrature couples to the cavity, and is measured by observing the cavity output. 

The spectra of the mechanical collective quadrature $X_+$ and its conjugate collective quadrature $P_+$ are given by \cite{supplement}
\begin{equation}
\label{eq:spectra}
\begin{split}
& S_{X_+}[\omega]  = \frac{1}{2} \LL( n^T_1 + n^T_2 + 1 \RR) S_0[\omega] \,, \\
& S_{P_+}[\omega] = \frac{1}{2}  \LL( n^T_1 + n^T_2 + 1+ 2n_{\m{BA}}  \RR) S_0 [\omega] \,,
 \end{split}
\end{equation}
where $S_0[\omega]=\sum_{\sigma=\pm}2\gamma/[\gamma^2+4(\omega+\sigma\Omega)^2]$ and 
\begin{equation}\label{eq:nBA}
n_{\m{BA}}= 2 C \frac{\kappa^2}{\kappa^2 + 4\Omega^2}(2n_c^T + 1)
\end{equation}
is the total measurement back-action, with $n_c^T$ being the thermal occupation of the cavity. The cooperativity is introduced as $C=4G^2/(\gamma\kappa)$ with $\gamma=(\gamma_1+\gamma_2)/2$. The back-action can be divided into classical and quantum contributions: $n_{\m{BA}}=n_{\m{BA,c}}+n_{\m{BA,q}}$ corresponding to, respectively, the terms $2n_c^T$ and $+1$ inside the parentheses in Eq.~\eqref{eq:nBA}. We define the occupation number by $n_{X_+}+\frac{1}{2}=\LL(\Delta X_{+}\RR)^2=(2\pi)^{-1}\int\!S_{X_+}[\omega]d\omega$, and similar for $n_{P_+}$. As is evident from Eq.~\eqref{eq:spectra}, only the $P_+$ quadrature is heated by back-action, while the measured quantity $X_+$ remains unaffected. Experimentally, there can be additional technical back-action via an increase in $n^T_i$ from equilibrium values due to pump current heating.

The mechanical spectrum $S_{X_+}[\omega]$ is faithfully reproduced in the cavity output spectrum $S_{\m{out}}[\omega]$. As shown in \fref{fig:setup}(b), it appears as sidebands at frequencies $\omega_a\pm|\Omega|$, on top of a noise floor due to experimental contributions \cite{supplement}. This allows for experimental reconstruction of the $X_+$ quadrature by a spectral analysis of the scattered pump microwave light.

\begin{figure}
\includegraphics[width=\columnwidth]{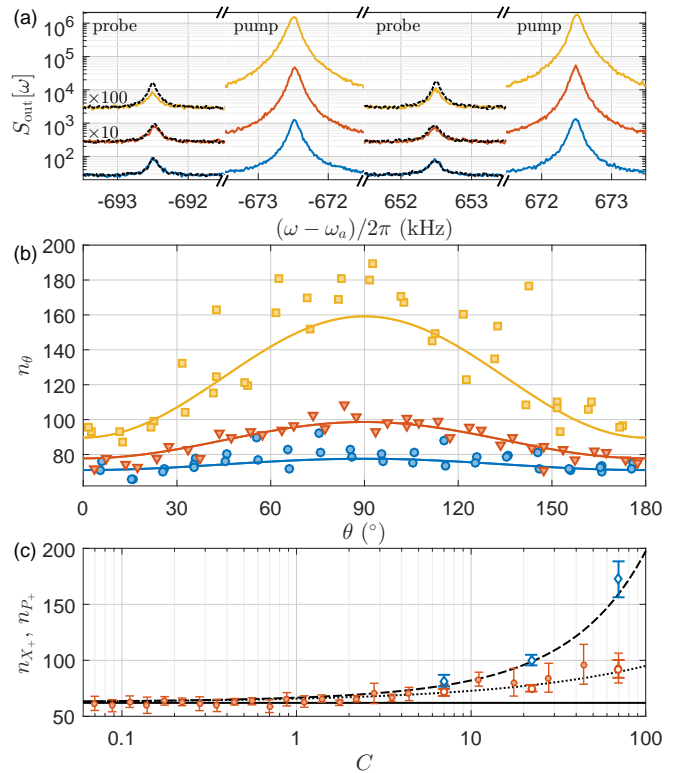}
\caption{\emph{Back-action tomography.} (a) Measured spectrum with pumps and weak probes for $C=7,22,70$ (bottom to top). Curves are offset by a factor 10 for clarity, as indicated. The tall peaks are due to the pumps, as was shown in \fref{fig:bae_power}(a). The short peaks are the probe signal, measuring $X_+$ (solid lines, $\theta=0^\circ$) and $P_+$ (dashed lines, $\theta=90^\circ$), respectively. (b) Effective occupation $n_\theta$ of the phase-dependent quadrature $X_{\theta}$ obtained from the probe signal. Data is shown for the same pump strengths as in (a). Solid lines show the modelled back-action and technical heating. (c) Effective occupations of $X_+$ (red circles) and $P_+$ (blue diamonds) as functions of measurement strength. Open symbols are extracted from the probe data in (b), while closed circles are measured from the pump spectra. Error bars show statistical spread of the data (95\% confidence) and thermal calibration uncertainty. The solid line shows the occupation expected from equilibration with the cryogenic environment, the dotted line incorporates technical heating, and the dashed line additionally includes the measurement back-action.}
\label{fig:bae_phase}
\end{figure}

We stress that the symmetry inherent in Eq.~\eqref{eq:hamilt} that protects $X_+$ from the back-action heating of $P_+$ requires closely matched single-photon optomechanical couplings. This thus introduces an additional experimental complication compared to the single-mode optomechanical BAE measurement. It should be noted that the coupling asymmetry cannot be compensated by tuning the pump power ratio. In practice, one needs to have $g_1\approx g_2$ within about 10\% accuracy, which is experimentally challenging. A theoretical description including deviations from the ideal case is thoroughly discussed in Ref. \cite{WoolleyBAE}.

As schematically depicted in \fref{fig:setup}(a), the mechanical oscillators \cite{Teufel2011b} are connected to opposite ends of a transmission line cavity resonator, nearly 1 mm apart from each other. The experiments are carried out in a dilution refrigerator at a temperature of 27 mK (unless stated otherwise). The cavity is probed using a transmission measurement. The cavity has the frequency $\omega_a\simeq2\pi\times5.5$~GHz, total linewidth $\kappa\simeq2\pi\times1.22$~MHz dominated by coupling to the output line with the rate $\kappa_\m{Eo}\simeq2\pi\times0.98$~MHz. The coupling at the input side line is $\kappa_\m{Ei}\simeq2\pi\times60$~kHz, and the internal decay rate is $\kappa_\m{I}\simeq2\pi\times180$~kHz. The mechanical oscillators have the  frequencies $\omega_1=2\pi\times10.0$~MHz and $\omega_2=2\pi\times11.3$~MHz, and linewidths, $\gamma_1=2\pi\times130$~Hz and $\gamma_2=2\pi\times 150$~Hz, respectively. For this sample we obtained $g_1/g_2\simeq0.94\pm0.02$, which allows for a nearly ideal two-mode BAE measurement~\cite{supplement}. The two pump tones need to be equal in amplitude as well, at the highest cooperativities presented here at 0.1\% accuracy. We can calibrate the pump amplitude ratio with about 5\% accuracy, and then use the pump power ratio as an adjustable parameter within the calibrated window. 

\fref{fig:bae_power} shows the BAE measurement results. In \fref{fig:bae_power}(a) the measured cavity output spectra are shown for different measurement strengths (cooperativities). \fref{fig:bae_power}(b) shows the integrated peak area, which corresponds to a photon flux \cite{supplement}
\begin{equation}\label{eq:naout}
n_{\m{out}} = \kappa_\m{Eo} C \frac{4\gamma\kappa}{\kappa^2 + 4\Omega^2}\LL(n_{X_+} + \frac{1}{2}\RR).
\end{equation}
For cooperativity up to $C\simeq5$, the data is in excellent agreement with the expected linear behaviour, indicating that the $X_+$ quadrature is not perturbed by measurement back-action. At the largest measurement strengths available in our experiment (limited by the technical requirement of equal pump tone amplitudes), a small increase in $n_{X_+}$ is observed. This is in good agreement with independently measured technical heating of oscillator 2, see \fref{fig:bae_power}(e). The measured $\gamma$, shown in \fref{fig:bae_power}(c), is independent of the measurement strength. As a comparison, we also carried out a measurement when only the red pump was switched on, characterized by a strong back-action. Under these conditions, the back-action is associated with increased damping of each mechanical oscillator by the amount $\Gamma_{\m{opt}}=4G^2\kappa/(\kappa^2+4\Omega^2)$, and consequently cooling of each down to an occupation $n_i=n_i^T\gamma_i/(\gamma_i+\Gamma_{\m{opt}})$ \cite{supplement}. As shown in \fref{fig:bae_power}(b)-(c), both the occupation and linewidth strongly deviate from those in the BAE scheme.

In order to calibrate the measurement results, we perform a BAE measurement while varying the cryostat temperature $T$. As shown in \fref{fig:bae_power}(d), for $T\gtrsim50~$mK the measured $n_{X+}$ is linear with $T$, indicating the system thermalizes with the environment. We use a linear fit to this data to calibrate the measurements of $n_{X+}$. At the operating temperature $T=27~$mK, the equilibrium occupation is $\frac{1}{2}\LL(n^T_1+n^T_2+1\RR)=62\pm2$. The output photon flux $n_{\m{out}}$ is subsequently calibrated by comparing to Eq. \eqref{eq:naout} for low power ($C\lesssim2$).

By its nature, the BAE measurement only accesses the unperturbed collective quadrature $X_+$. To fully quantify the measurement back-action, we perform a second experiment, where we additionally apply a second, weak BAE measurement. It is realized by two \emph{probe} tones, offset $\delta=2\pi\times20~$kHz below the pump tones, as shown in \fref{fig:setup}(a). The probes perform a weak measurement, with cooperativity $C_\m{probe}\approx 0.3\ll C$, such that the probe itself causes negligible back-action. All tones are phase-locked to a common reference. By adjusting the phase of one of the probe tones by an amount $2\theta$, the probes measure the generalized collective quadrature $X_+^{\theta}=X_+\cos\theta+P_+\sin\theta$, as depicted in figure \fref{fig:setup}(d). The probe tones cause two additional peaks in the output spectrum, whose area corresponds to the phase-dependent occupancy $n_{\theta}$ as in Eq.~\eqref{eq:naout}. Here, $n_{\theta}=n_{X_+}\cos^2\theta+n_{P_+}\sin^2\theta$, assuming the correlations between $X_+$ and $P_+$ are negligible. Note that other tomographic techniques for two-mode mechanical systems have recently been demonstrated \cite{YamaguchiSqu2014,Vengalattore2015TwoSqu,Marin2016TwoSqu}.

In \fref{fig:bae_phase}(a) we display the total spectrum in the complete pump-probe configuration. It consists of four peaks, two corresponding to the strong pumps measuring $X_+$ and two corresponding to the weak probes measuring $X_+^{\theta}$. In \fref{fig:bae_phase}(b) we plot the measured occupancy $n_{\theta}$, showing the strong phase-dependence of the measurement back-action at large measurement strengths. \fref{fig:bae_phase}(c) shows the quadrature occupations $n_{X_+}$ and $n_{P_+}$, measured from a sinusoidal fit to the data in panel (b), as well as $n_{X_+}$ measured by the pumps, as a function of pump power. The data is well-described by a theoretical prediction which includes quantum back-action on $P_+$ and technical heating of $n_2^T$. The quantum back-action (up to $n_{\m{BA,q}}\approx 63$) dominates over the classical contribution ($n_{\m{BA,c}}\approx 6$). The technical heating was independently calibrated with standard optomechanical cooling measurements, shown in \fref{fig:setup}(e)-(f), using a single pump with frequency $\omega_a-\omega_i$ ($i = 1,2$). For the two-tone measurement the heating is given by the total pump power. We typically observe heating in varying amounts in different samples~\cite{Squeeze,multimode2012}, likely related to the presence of surface two-level systems~\cite{simmonds04}.

\begin{figure}
\includegraphics[width=\columnwidth]{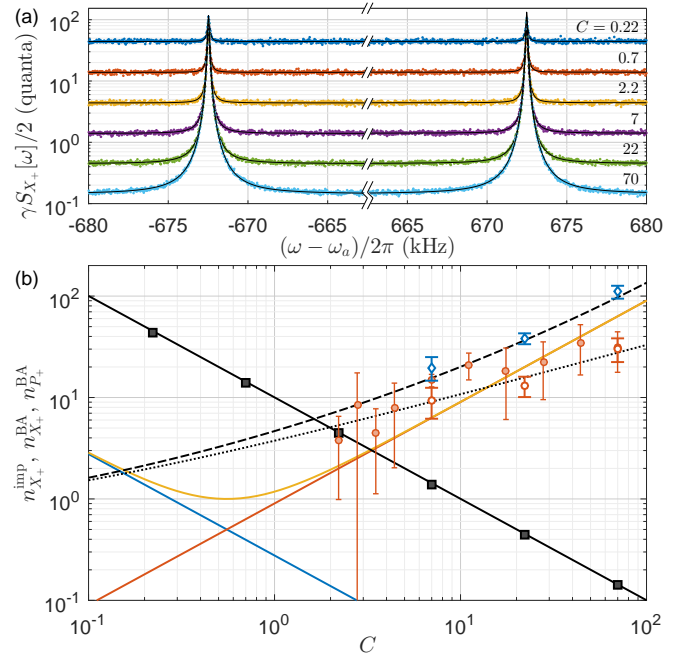}
\caption{\emph{Mechanical spectra and measurement precision.} (a) Measured mechanical spectra around $\omega_a\pm|\Omega|$ with different cooperativities as marked. (b) Measurement imprecision $n_{X_+}^\mathrm{imp}$ (black squares), technical back-action $n_{X_+}^\mathrm{BA}$ on $X_+$ (red circles), and conjugate back-action $n_{P_+}^\mathrm{BA}$ on $P_+$ (blue diamonds). Symbols as in \fref{fig:bae_phase}(c). Colored lines show quantum-limited imprecision (blue), quantum back-action (red), and their sum (yellow), the minimum of which is the SQL. Dotted (dashed) lines show the modelled $n_{X_+}^\mathrm{BA}$ ($n_{P_+}^\mathrm{BA}$) including technical heating. A fit to the imprecision (black solid line) corresponds to $n_\mathrm{amp}=28$.}
\label{fig:SQL}
\end{figure}

An important benchmark for a sensitive measurement is that the total measurement imprecision is below the fundamental quantum limit for continuous position detection. This has been demonstrated in single-oscillator optomechanical systems \cite{Lehnert2009SQL,Kippenberg2010SQL,Khalili2012SQL,Kippenberg2015FB}. The effective quadrature occupation can be written as $n_{X_+}=n_{X_+}^{T}+n_{X_+}^\m{BA}+n_{X_+}^\m{imp}$, where $n_{X_+}^{T} = (n^T_1 + n^T_2)/2$ is the initial occupation to be measured, $n_{X_+}^\m{BA}$ is the residual back-action including technical heating, and $n_{X_+}^\m{imp}$ is the imprecision noise. The latter, assuming the cavity output is measured with a high-gain phase-insensitive amplifier, is \cite{supplement}
\begin{equation}
n_{X_+}^\m{imp} = \frac{1}{C}n_c^T + \frac{1}{8C}\frac{\kappa^2+4\Omega^2}{\kappa\kappa_\m{Eo}}\LL(n_\m{amp}+1\RR),
\end{equation}
where $n_{\m{amp}}$ is the noise of the amplifier. The imprecision $n_{X_+}^\m{imp}$ can be made arbitrarily small by increasing the cooperativity. In the case of $n_c^T=n_\m{amp}=0$, the back-action noise on $P_+$ and imprecision noise of $X_+$ satisfy $n_\m{BA}n_{X_+}^\m{imp}=1/4$, which arises due to phase-insensitive amplification in the measurement chain. This relation is of a similar form as the Heisenberg imprecision back-action uncertainty relation for continuous position detection \cite{GirvinReview}. Notice that in the BAE case, however, the back-action only affects the unmeasured quadrature (which is dynamically decoupled from the measured quadrature) and hence has no effect on the spectrum of the measured quantity $X_+$. 

\fref{fig:SQL}(a) shows sample measured spectra $S_{X_+}[\omega]$, scaled such that the peak height corresponds to the number of quanta. For increasing cooperativity the imprecision (noise floor) is reduced. In \fref{fig:SQL}(b) we show the measurement imprecision and back-action against cooperativity. The back-action is calculated from the data in \fref{fig:bae_phase}(c) by subtracting the initial thermal occupation $n_{X_+}^{T}$, and shown for $C > 2$ where the signal-to-noise ratio is larger than one. At large measurement strengths, the measurement imprecision is well below the quantum zero-point fluctuations of the oscillators. Furthermore, the total back-action and imprecision of $X_+$ is below the quantum back-action by $3.2\pm1.0$ dB, demonstrating that our system out-performs a perfect phase-insensitive position measurement. The stated uncertainty is dominated by the statistical uncertainty (95\% confidence), calculated from the spread of the data in \fref{fig:bae_phase}(b), but also includes the calibration uncertainty of $\pm 0.3$~dB.

In summary, we have performed sensitive measurements of the collective motion of two mechanical oscillators without disturbance from the quantum back-action of the measurement. The measurement sensitivity exceeds the standard quantum limit, leading to a strong back-action observed in the canonically conjugate observable. The residual back-action is dominated by technical heating of one mechanical oscillator, which does not correspond to a fundamental limitation. Neglecting technical heating while keeping other parameters unchanged, the back-action would reduce to a few quanta, closely approaching the SQL. Further, with an improved imprecision $n_\mathrm{amp}\sim1$, this can be reached at small pump power, $C\sim1$. The two-mode back-action evading measurement can be used for quantum state tomography and for measuring both quadratures of a force without a quantum limit. This work also provides a foundation for the preparation and detection of macroscopic mechanical entanglement.

\begin{acknowledgments}
We would like to thank F. Massel and T. T. Heikkil\"a for useful discussions. This work was supported by the Academy of Finland (contract 250280, CoE LTQ, 275245) and by the European Research Council (615755-CAVITYQPD). The work benefited from the facilities at the Micronova Nanofabrication Center and at the Low Temperature Laboratory infrastructure.
\end{acknowledgments}

\end{document}